# Applying Test Paradigms in a Generic Tutoring System Concept for Web-based Learning

Matthias R. Brust
University of Trier
System Software and Distributed Systems
Department of Computer Science
D-54286 Trier
Germany
Phone: +49 651 202 3313
Fax: +49 651 201 3842
Email: brust777.geo@yahoo.com

*Abstract* Realizing test scenarios through a tutoring system involve questions about architecture and didactical methods in such a system. Observing the fact that traditional tutoring systems normally are domain-static, this paper shows investigations for a generic domain-independent tutoring system for utilizing test scenarios in computer-based and web-based environments. Furthermore, test paradigms are analyzed and it is presented an approach for realizing functionality for applying test paradigms in the presented generic tutoring system architecture by an XML-specified language.

I. INTRODUCTION

There is an understandable desire to use computers in education, partly to aid illustrations of complex phenomena and partly to improve efficiency. Tutoring Systems are computational environments to support learning [Excell 2000].

The initial idea was to make computer science tests (discipline "Operating Systems") by a tutoring system based on test simulation and knowledge handling.

The investigations, however, showed that there are many domain-specific and domain-dependant tutoring systems. Thus, it didn't seem interesting from a scientific point of view to add one more to the list, but to develop a generic tutoring system which could be used for any domain and for Computer Science tests especially. Test simulation in a tutoring system involves questions of how the professor creates tests. Consequently, this paper trades the following questions:

- How should a tutoring system be designed to be generic?
- What advantages and disadvantages does a generic tutoring system have in comparison to a traditional domain-bounded tutoring system?
- What kinds of test paradigms exist or are desirable in such a system dedicated for test simulating and knowledge handling?
- How can a generic Tutoring System offer functionality to apply test paradigms?

The rest of this paper is organized as follows. Section II describes the requirements for a generic tutoring system concept. Section III and IV explain the theoretical and practical investigations of the test. Practical issues are discussed in section V. Conclusion and future work can be found in section VI.

II. A GENERIC TUTORING SYSTEM CONCEPT

The traditional Tutoring System architecture is developed for a specified domain.as illustrated in Picture-1.

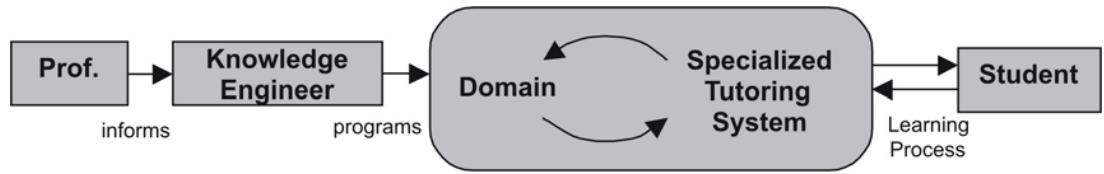
Picture-1 – Developing and learning process in traditional tutoring system architecture.

In consequence of this traditional architecture, this work spans several persons within an educational setting and traditional tutoring systems are extremely domain-dependant and often neither methods nor knowledge can be reused [Devedzic et al. 1998]. In addition, the knowledge engineer must have programming skills in order to enhance and to evaluate the system.

In the proposed generic tutoring system concept the domain was separated from the tutoring system. Picture –2 shows the proposed architecture.

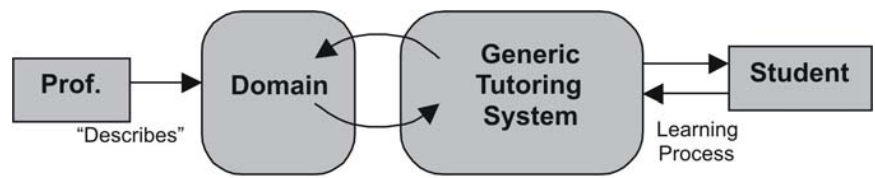
Picture 2 – Developing and learning Process in a generic Tutoring System architecture.

The main benefit of such separation is the possibility to apply an XML-specified language for tests and, thus, to establish standard directives for system design [Leiva et al. 2000]. Based on such designed language, the professor will be able to model a test without deep programming skills and a tutoring system-independent domain design is possible.

III. TEST PARADIGMS AND EVALUATION (DESIGN OF TESTS)

Student evaluation in higher education is normally based on written tests. In special settings (i.e. University of Trier-Germany) in which the amount of students in a Computer Science discipline barely reaches 15 individuals, we may expect more efficient and personalized interaction, such as oral tests.

Conversely to written tests, an oral test offers different aspects among which evaluation may occur.

Written Test Paradigm

The typical format for traditional tests consists of an initial section, a section of a practical problem and a residual small proportion of marks are reserved for a small test of advanced abilities in order to target those who aim for first-class marks.

This test paradigm tries to map the student's current state of knowledge and some other abilities that the professor wants to know [Remmers et al. 1959].

Oral Test Paradigm

Analyzing professors test behavior in Computer Science courses at the University of Trier based on student-made protocols after the test, it was possible to identify mainly three oral test paradigms that can described as follows:

1. The professor chooses a more advanced (right answer) or a more basic (wrong answer) question. Evaluation occurs by considering the right and wrong answers gotten during the test.
2. The professor elaborates a problem situation aiming to find out acquired learning processes, scientific abilities and how the student correlates knowledge. The main assumption here is that the student acquires knowledge by "on-line" correction, which will enable the student to attack advanced questions.

3. The professor presets the order of the questions and can reconfigure it to tolerate slips, faults, and errors until a certain limit. If the limit is crossed, the professor can repeat the last N questions or use this crossing to better estimate the evaluation.

In fact, evaluation praxis mixes these test paradigms and therefore the analyzed oral test protocols mainly contain all the four described test paradigms.

IV. IMPLEMENTATION OF FUNCTIONALITY FOR APPLYING TEST-PARADIGMS

The meta-language XML was used as principal technology for implementing functionality for applying test paradigms described above. The implementation occurs in the generic Tutoring System as described in Section II.

First, it was necessary to conceptualize these test paradigms. This resulted in descriptions of the main question: What question has to be presented to the student afterwards? These descriptions are in the terms of tutoring systems types of selections that correspond to each test paradigm described above.

1. Free Selection
2. Causal Link-oriented Selection
3. (Dynamic) Ordering Constrain-oriented Selection
4. Balanced Constrain-oriented Selection

The *free Selection* leads to questions in the order they appear in the XML-File and represents the described written test paradigm.

*Causal Links-oriented Selection* is sensitive to right and wrong answers. The attribute for the case, in which the answer was right, is called *forward* and the attribute for the wrong answer *backward*, respectively. Both attributes are specified as IDREFS-type (table 1).

| Causal Links | ```
...
<xTest id="A">
    <Print>...</Print>
    <Right forward="B" backward="A">...</Right>
        ...
    <Wrong backward="A">...</Wrong>
</xTest>
<xTest id="B">
    <Print>...</Print>
    <Right forward="C" backward="A">...</Right>
        ...
    <Wrong forward="D" backward="B">...</Wrong>
</xTest>
<xTest id="C">
    ...
</xTest>
...
``` |
|---|---|

Table-1 – An instance of causal links in XML

Causal Links can create cyclic structures by references. This has to be dealt with on the application side and not on the parser side – how it would be desirable. XML by its own does not offer an adequate mechanism to handle it. Multiple-Choice-Questions, for instance, offers more than one alternative answer. Thus, the system designer has to choose a meaningful and consistent interpretation [Brust 2002].

*(Dynamic) Ordering Constrain-oriented Selection* follows determined order for selecting questions. The important attribute is called *order* (type IDREFS) with possible values *normal* or *forced*. A forced ordering constrain-question can just be called by other question by reference. This construction takes the existence of a question that are subparts of other questions and that does not make sense alone contextually into account.

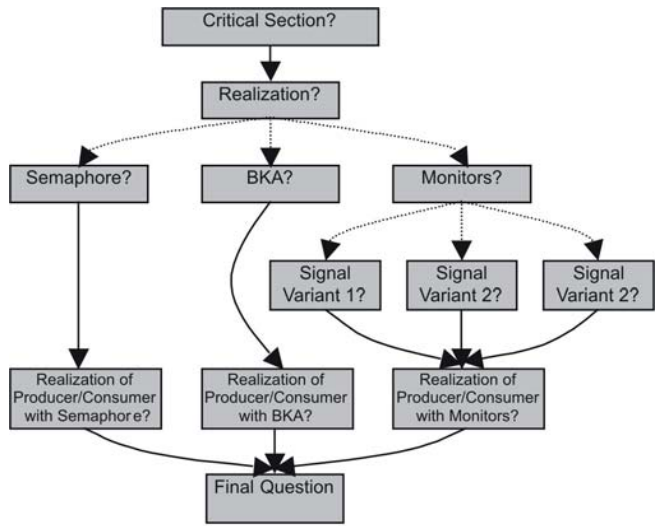

Picture-3 – Example for realizing an "operating system" test schematically

Another attribute called *orderType* was introduced to enable the system to choose randomly different constrains on multi-using of one test. The attribute *orderType* may have the values *normal* or *alternative*. For illustration see Picture-3 and Table-2. The dotted lines denote the alternative constrains.

| | |
|---|---|
| *Example of a dynamic ordering constrains in combination with balanced ordering constrains* | ```
...
<Test balanced="3 70">
    <xTest id="CriticalSection" order="Realization FinalQuestion">
        ...
    <xTest id="Realization" order="Semaphore BKA Monitor"
                            orderType="alternative" type="forced">
        ...
    <xTest id="Semaphore" order="ProdConsSemaphor" type="forced">
        ...
    <xTest id="ProdConsSemaphor" type="forced">
        ...
    <xTest id="BKA" order="ProdConsBKA" type="forced">
        ...
    <xTest id="ProdConsBKA" type="forced">
        ...
    <xTest id="Monitor" order="Signal1 Signal2 Signal3"
                        orderType="alternative" type="forced">
        ...
    <xTest id="Signal1" order="ProdConsMonitor" type="forced">
        ...
    <xTest id="Signal2" order="ProdConsMonitor" type="forced">
        ...
    <xTest id="Signal3" order="ProdConsMonitor" type="forced">
        ...
    <xTest id="ProdConsMonitor" type="forced">
        ...
    <xTest id="FinalQuestion" type="forced">
        ...
</Test>
``` |

Picture-3 – Realizing schematic test in Picture-3 by the proposed XML-specified language

*Balanced constrain-oriented selection* acts superior from a conceptual point of view than the forgoing selection types. Interestingly, this hierarchal attribute could be transmitted in the language design.

The configuration is done in the root-Element *Test*. The attribute balance may have the values **n** and **p**. The **n** value implies an arithmetic average **a** for the last **n** selected questions. The **p** value determines if the average **a** is greater than **p** (see Table-2). In this case, the system may continue selecting the next question and following the ordering constrains, causal links or free selection. In the other case the selection repeats the last **n** questions.

## V. PRACTICAL ISSUES

The tutoring system's automatic planning component needs a fundamental over which it selects questions [Collins 1996]. Different types of selections were presented in the forgoing section. Practically, it rises up the questions: When is a test finished? How to treat combined selection-types?

Normally, tests should be considered as finished, if all questions were answered. Based on dependence caused by causal links, (dynamic) ordering constrains, and balanced constrains, a test may also be considered finished, when all final-questions (questions that are the last in the dependency line) were reached or all questions in one test were correctly answered [Brust 2002].

Question selection can depend on how often a question was referenced (*Eventing*) or it can depend on the order it was referenced through the test [Brust 2002].

The used implementation demonstrates two sets of selection. The first set is a priority queue with question-references from causal links, ordering constrains, and balanced constrains. The second set contains all questions that can be called initially or when there is no element in the priority queue. If the two sets do not contain any element, the test is considered as finished.

Ordering constrain-referenced questions are inserted in the priority queue with higher priority than causal link-referenced questions. This sanction is necessary because ordering constrain-referenced questions generally have a context-sensitive logical order and the selection of causal link-referenced questions would interrupt and interfere this order educationally.

The applied XML-specified language includes elements to describe common question formats such as Multiple-Choice-Questions, True-False-Questions, Completion-Questions, Numeric-Questions, etc [Brust 2002]. Moreover, the knowledge engineer may enhance the system by adding new DTD (Document Type Definition) specified components.

## IV. CONCLUSION AND FUTURE WORK

Using an XML-specified language tore down four main deficiencies in tutoring system design. (1). An XML-specified language for tests in web-based learning environments offers a standard for test design, (2) based on this specification, the professor is able to model or design a test without deeper programming knowledge. For this reason, the eficiency of specialized methods for accessing domain-knowledge was given up. But for test dedicated tutoring system, it is sufficient how test realizations based on test protocols in the described system showed. (3) A tutoring system-independent domain design is possible and (4) to realize a standardized specification for tests in web-based learning environments. In more sophisticated cases, like applying test paradigm, it would be helpful to access standardized reference-types in XML to avoid cyclic structures through references.

Practically, it was very simple to create functionality in a generic tutoring system for applying different test paradigms with XML. It also raised conflicts by allowing simultaneous test paradigms, especially when ordering constrain loses their priority by interaction with balanced constrain. Here it should be found a good tradeoff between technical and educational points of view.